# The Programmable Liquid-crystal Active Coronagraphic Imager for the 4-m DAG telescope (PLACID) instrument: installation and commissioning update

Jonas G. Kühn,*[a] Ruben Tandon,[a] Lucas Marquis,[a] Liurong Lin,[a] Derya Öztürk Çetni,[b] Iljadin Manurung,[a] Axel Potier,[a] Laurent Jolissaint,[c] Audrey Baur,[c] Daniele Piazza,[a] Mathias Brändli,[a] Martin Rieder [a]

[a]Division of Space and Planetary Sciences, University of Bern, Sidlerstrasse 5, 3012 Bern, Switzerland; [b]Turkiye National Observatories, Atatürk University Astrophysics Research and Application Center ATASAM, Atatürk University Campus, 25240 Yakutiye/Erzurum, Turkey
[c]University of Applied Sciences HEIG-VD, Route de Cheseaux 1,
1401 Yverdon-les-Bains, Switzerland;

## ABSTRACT

The Programmable Liquid-crystal Active Coronagraphic Imager for the DAG telescope (PLACID) instrument is a novel high-contrast direct imaging facility that was recently installed on the new Turkish 4-m DAG telescope. In brief, PLACID consists in a fore-optics coronagraphic intermediate stage platform, installed in-between the TROIA XAO system and the DIRAC HAWAII-1RG focal-plane array. The PLACID instrument was delivered to ATASAM campus facilities in March of 2024, and transported to summit in October of 2024. In February of 2025, the PLACID optical breadboard was craned to the DAG observatory floor, and successfully installed on the optical table of the diffraction-limited Nasmyth platform of the 4-m telescope. Following the official DAG Acceptance milestone in the spring of 2025, Assembly, Integration and Validation (AIV) activities have started in July of 2025, when PLACID was cabled up with all active components fully interfaced and tested for functional integrity. PLACID optical commissioning with the internal calibration source is expected to take place in October of 2025, once the TROIA XAO system optically aligned. When on-sky by early 2026, PLACID will be the world's first "active coronagraph" system, fielding a customized spatial light modulator (SLM) acting as a dynamically programmable focal-plane phase mask (FPM) coronagraph from H- to Ks-band. This will provide a wealth of novel options to observers, among which software-only abilities to change or re-align the FPM pattern in function of observational conditions or science requirements. Future features will include non-common path aberrations (NCPA) self-calibration, angular differential imaging (ADI) coronagraphy for binary or triple stars, as well as coherent differential imaging (CDI). We hereby present the PLACID AIV activities that have taken place over the last twelve months, and the next steps for commissioning the instrument internally, and on-sky later this year.

## 1. INTRODUCTION

Direct "high-contrast" imaging (HCI) of exoplanets allows for in-situ observing of exoplanets in their host stellar environment, enabling to test planet formation theories and study disk interaction processes, but also spectroscopic characterization of the atmospheric composition of exoplanets. However, the number of confirmed detections of planetary mass objects with HCI has not exceeded a few dozen (see e.g. [1-4]), despite several generations of dedicated high-contrast instruments deployed on 8-m class telescopes [5,6]. Detecting exoplanets with HCI from the ground is admittedly a challenging task, because of the extreme contrast ratio ($10^{-4}$ to $10^{-10}$) required at angular separations as small as a few diffraction beam widths ($\lambda/D$ units). The upcoming class of extremely large telescopes (ELTs), fielding 25- to 40-m diameter primary mirrors, will ease some of these challenges, notably by improving sensitivity and angular resolution, but several of the aforementioned issues affecting direct imaging and coronagraphy will remain, or may even worsen. Those include residual differential atmospheric refraction, resolved nearby giant stars (a side-effect of using bigger telescopes), non-common path aberrations (NCPAs), pupil registration stability, and non-ideal segmented telescope apertures.

*jonas.kuehn@unibe.ch

The first idea of "adaptive coronagraphy" was originally proposed by Bourget et al. in 2012, using a compressed mercury (Hg) drop to generate a Lyot coronagraph with tunable diameter occulting spot [7]. In 2016, we proposed [8] to use liquid-crystal on-silicon spatial light modulators (LCOS SLMs, see e.g. [9]) as active programmable focal-plane phase masks (FPM) coronagraphs, taking advantage of their resolution often exceeding 1 Mpixel and ~10 μm pixel pitch, allowing for sufficient sampling of the telescope point-spread function (PSF) in the coronagraphic focal-plane. On the negative side, SLMs imprint a scalar phase retardance, hence limiting the broadband contrast performance as compared to e.g. vectorial phase plates [10,11]. Another drawback is that classical SLMs require linearly-polarized input light, thus limiting optical transmission to less than 50% for unpolarized light. Nevertheless, those limitations may not represent a contrast bottleneck from the ground, where post-AO wavefront errors (WFE) residuals and leakage from the central obstruction caused by the telescope secondary mirror dominate the leakage budget [12]. On the positive side, such a SLM-based programmable FPM coronagraph will enable the astronomers to adapt to observing conditions (seeing, wind profile, etc.) in real-time, by freely selecting an optimal FPM in a trade-space between inner-working angle (IWA) and robustness to low-order aberrations, in particular tip/tilt jitter. Observers will also be able to adapt to the specifics of the science target, for example between a follow-up strategy or a blind survey, resolved or unresolved stars (to be relevant in the ELTs era), or in particular the multiplicity of the target. In the latest scenario, a programmable focal-plane coronagraph can be setup to null several stars in the field-of-view (FOV) of the instrument, enabling niche science case for high-contrast observations around challenging compact binary or triple stars systems of similar magnitudes [8,13]. As SLM-based approaches enable to change the FPM pattern by software at a rate of 30 Hz or more, a multiple-star HCI observing mode can be developed that allows for pupil-tracking operation, hence enabling Angular Differential Imaging [14] of binary or triple stars, with the SLM-generated FPM pattern following the sky rotation. Extra potentially interesting features of adaptive FPM coronagraphy include self-calibration of NCPAs, with the SLM generating a Zernike wavefront sensor (WFS) pattern [15], but also its advantageous phase-shifted variants [16,17], without any other physical change in the optical path. Finally, the SLMs can be used to introduce focal-plane phase diversity at specific modulation frequencies, potentially enabling time-domain coherent differential imaging (CDI), in order to dynamically disentangle coherent speckles from bona-fide incoherent off-axis astrophysical sources [17].

## 2. THE PLACID INSTRUMENT FOR THE DAG TELESCOPE

The Programmable Liquid-crystal Active Coronagraphic Imager for the 4-m DAG telescope (PLACID) instrument project is led by a consortium of Swiss universities consisting of the University of Bern and the University of Applied Sciences HEIG-VD, under procurement from the Atatürk University Astrophysics Research and Application Center (ATASAM). PLACID is the high-contrast exoplanet direct imaging instrument for the new Eastern Anatolia Turkish National Observatory (DAG), a 4-m Ritchey-Chrétien telescope, located atop Karakaya Ridge (3,100-m ASL) in the Erzurum province [18]. The PLACID contract was awarded in October 2020 and passed the Final Design Review (FDR) milestone in December 2021. The instrument itself was assembled in HEIG-VD laboratory facilities in Yverdon, Switzerland, where it passed Factory Acceptance (FA) a year later in December 2022, followed by a successful Delivery Readiness Review (DRR) in August 2023. Finally, PLACID was shipped by air and road transport to ATASAM campus facilities in Erzurum, Turkey, in February of 2024, followed by an on-site delivery inspection review from the PLACID team in March of 2024.

The retained optical design for PLACID is shown in Figure 1-Left. PLACID consists in a fore-optics intermediate coronagraphic beam train, fielding a customized H-/Ks-band 1152x1920 SLM from Meadowlark Corp. in a coronagraphic reflective focal-plane (off-axis angle ~ 5°). In practice, a pair of fold-mirrors (denoted PUM1 and PBM3 on Fig.1) can be lifted up to intercept the beam coming from the TROIA XAO system [19] towards the DIRAC H1RG detector [20], in order to enable the PLACID active coronagraphic imaging mode (as opposed to classical non-coronagraphic AO imaging mode). In this configuration, PLACID will slow down the telescope AO beam from F/17 to F/60, in order to provide a PSF spatial sampling of 10 SLM pixels per λ/D (at H-band) in an intermediate F/60 reflective coronagraphic focal-plane, where the SLM panel is located. In combination with a double-pass wire-grid linear polarizer (LP) located right in front of the SLM (not shown in Fig.1-Left), the LCOS SLM panel will act as programmable coronagraphic FPM and diffract the on-axis stellar light, which can be blocked out downstream in the post-coronagraphic Lyot pupil-plane (LPP, see Fig.1-Left), where a motorized filter wheel is located with a variety of Lyot masks (see [21] for technical details). The remaining PLACID optics are designed to reconfigure the output beam at F/17 before feeding the DIRAC detector. A general overview of the NIR XAO-assisted diffraction-limited Nasmyth platform instrumentation configuration of the DAG observatory is presented in Figure 1-Right. More technical details on the instrument, its specifications and Factory

Acceptance performance, as well as a brief description of its Python-based graphical user interface (GUI) are provided in Kühn et al. 2024 [21].

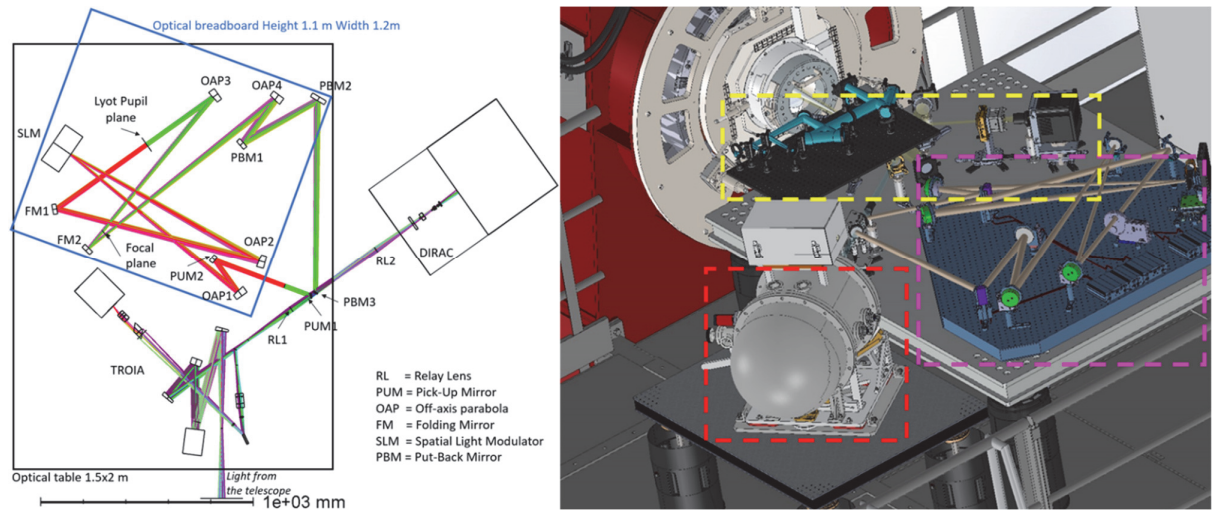

Figure 1. (Left) The PLACID Optical Zemax Design. The pair of mirrors PUM-1 and PBM3 is on a motorized vertical stage, in order to enable or bypass the PLACID beam path in-between the TROIA XAO system and the DIRAC H1RG detector (choice of coronagraphic or non-coronagraphic imaging mode); (Right) 3-D CAD representation of the DAG AO Nasmyth platform instrumentation configuration. The PLACID breadboard is outlined with the purple dashed rectangle, while the TROIA XAO and DIRAC HAWAII-1RG detector are outlined with yellow, respectively red, dashed rectangles.

## 3. STATUS UPDATE AND OVERVIEW OF ONGOING AIV ACTIVITIES

### 3.1 PLACID transport to Karakaya Ridge summit and DAG facilities (November 2024)

As previously mentioned, the PLACID instrument was shipped from Switzerland to Erzurum, Turkey, on February 23$^{rd}$ 2024, and delivered to the ATASAM facilities on the Erzurum Atatürk University campus on March 8$^{th}$, 2024. The Atatürk University campus is located close to downtown Erzurum, at an altitude of about 2,000-m and within 2-hours driving distance from the DAG observatory summit. The PLACID team inspected the shipment integrity on-site in mid-March 2024, but the DAG telescope commissioning activities involving ATASAM and their contractors AMOS and EIE were still ongoing at the summit through most of 2024. Hence PLACID remained in storage conditions in ATASAM campus facilities for about 6 months, waiting for the green light from ATASAM to be transported to the summit site.

The road condition to the Karakaya Ridge summit of the DAG observatory are practical for large transport trucks roughly only from the May to November period. Indeed, the presence of snow – or melting snow combined with mud – prohibits using other means of transport than snow cats or snow mobiles, given that the access road is unpaved from the Konakli ski resort (approx. 2,400-m elevation) to the last few hundred meters before the summit at 3,200-m. In the fall of 2024, most large commissioning activities related to the DAG observatory were completed at the site, but the diffracted-limited Nasmyth platform thermal enclosure – controlling humidity and temperature (10-20°C) levels – was still in the process of being installed (see Figure 2 for details on the Nasmyth infrastructure). Hence it was decided to preemptively move the PLACID instrument in its shipping configuration, along with the TROIA XAO facility, to the observatory site ahead of the snow season. The truck transport of both PLACID and TROIA from Erzurum ATASAM facilities to the DAG site took place on November 5$^{th}$ 2024, right during the first light snow falls (see Figure 3). The shipping crates for both instruments were placed in a temperate storage room within the observatory main building, right next to the vertical shaft leading to the dome floor (Fig.3).

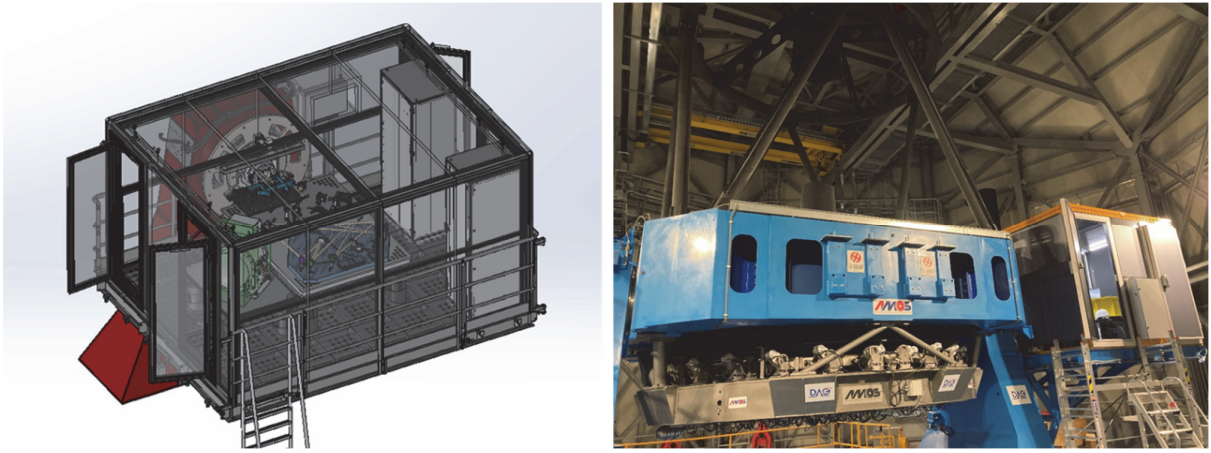

Figure 2. (Left) CAD design for the diffraction-limited Nasmyth platform of the DAG telescope, including the thermal enclosure (rendered as transparent here, but opaque in reality); (Right) Actual on-site configuration, with the access ladder.

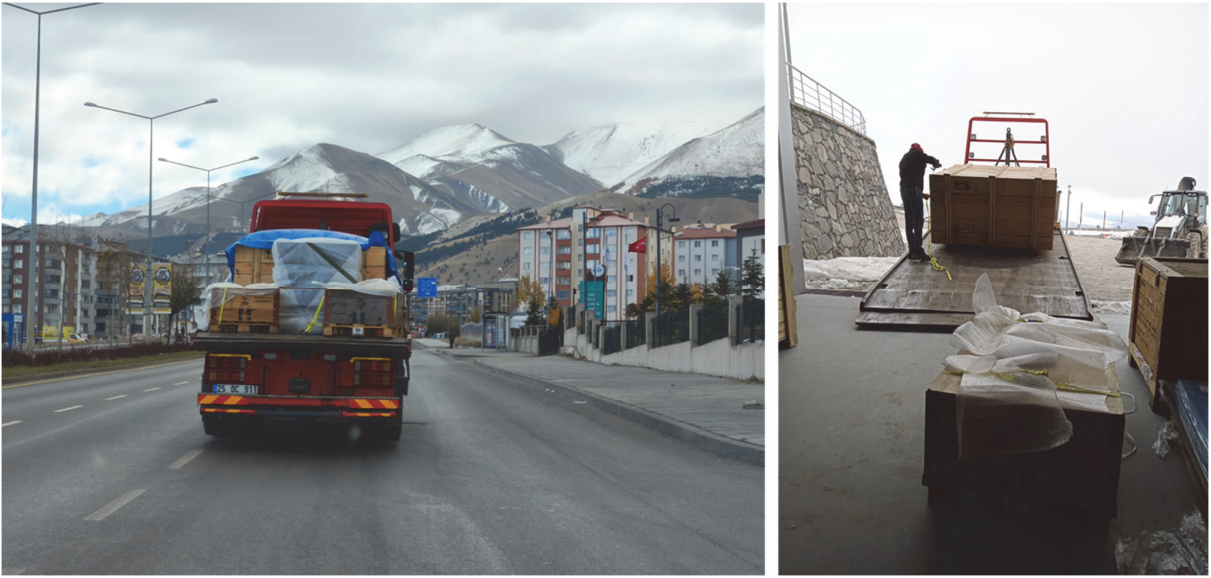

Figure 3. Both the PLACID exoplanet imager instrument and the TROIA XAO system were transported from ATASAM facilities on Atatürk University campus, in downtown Erzurum, to the DAG summit on November 5$^{th}$, 2024.

**3.2 PLACID physical installation on diffraction-limited Nasmyth platform (February 2025)**

The thermal enclosure of the diffraction-limited Nasmyth platform of the DAG telescope (Fig.2) was successfully installed over the new year period leading to 2025, as well as the vibration-insulated optical tables, hence the telescope operator ATASAM provided the go-ahead for starting the installation of the Nasmyth instrument suite in early January of 2025. We here recall that the instrument suite to be installed on the diffraction-limited Nasmyth of the DAG telescope comprises (in the order of the beam train propagating): the KORAY K-mirror derotator [22], the DAGOS internal calibration source [18], the TROIA XAO system fielding a Pyramid wavefront sensor and a 468-actuators deformable mirror [19], the (optional) PLACID coronagraphic stage [21], and the DIRAC HAWAII 1RG infrared array [20]. The latter has its own optical table (see Figure 2), and is scheduled for a later installation in February of 2026, but DAGOS, TROIA and PLACID share the same 1.1 x 1.2 m$^2$ Newport optical table, while KORAY is installed inside the telescope flange. Although it is located downstream of DAGOS and TROIA, the PLACID instrument is mounted on a dedicated 65-mm thick honeycomb breadboard (see Fig.1-Right). Therefore the PLACID breadboard and the opto-mechanical components atop represent the bulk of the mass – about 140 kg – to be installed on the main optical table. Hence the HEIG-VD team in charge of the KORAY and TROIA commissioning required the bulk of the PLACID mass to be installed on the optical table prior to their alignment work.

The ATASAM and PLACID teams performed the physical installation of the PLACID main breadboard on January 31$^{st}$ 2025. To this end, the PLACID breadboard was detached from its shipping enclosure and lifted to the dome floor with the DAG heavy duty crane, through the main shaft of the observatory (see Figure 4). For the purpose of this procedure, the top panel (ceiling) of the diffraction-limited Nasmyth platform was temporarily removed, and the PLACID instrument breadboard was successfully craned to its final location atop the Nasmyth main optical table (see Figure 5). A protective cover (not shown) was setup atop the PLACID breadboard by the Uni Bern team, at to the end of this campaign.

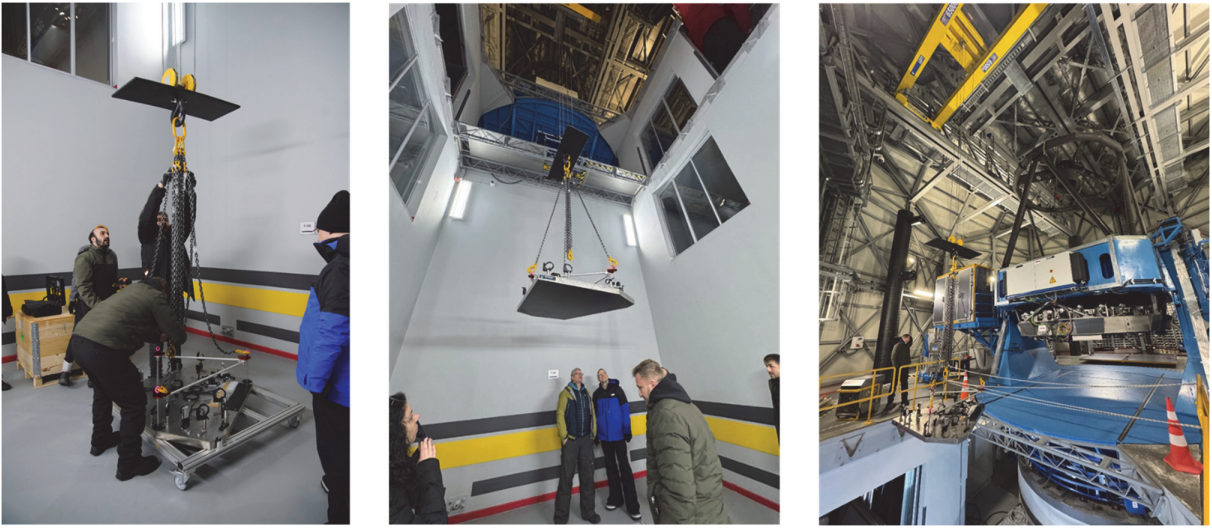

Figure 4. The PLACID breadboard was detached from its shipping enclosure, and then craned from the observatory storage bay to the dome floor through the main shaft on January 31, 2025.

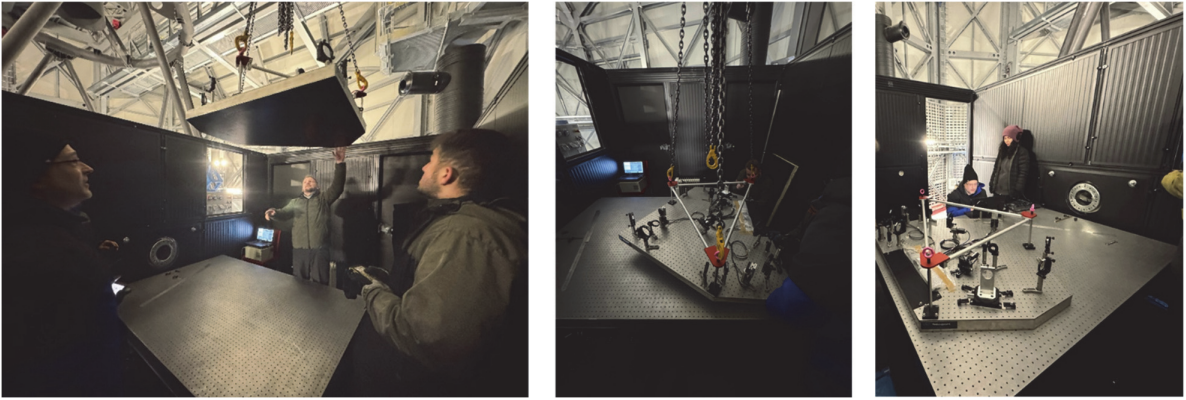

Figure 5. The PLACID breadboard – with the optics and controllers unmounted – was installed atop the main optical table of the diffraction-limited Nasmyth platform of the DAG telescope on January 31, 2025. The ceiling panel of the thermal enclosure was temporarily removed to this purpose.

### 3.3 PLACID cabling and interfacing campaign (July 2025)

The time period spanning March to June 2025 did not see much progress regarding the DAG Nasmyth instrumentation suite, sadly enough, mostly caused by delays associated to passing the (preliminary) Acceptance (PA) milestone of the main observatory. The PA milestone of the telescope was officially passed between the Turkish operator ATASAM and the main DAG contractor AMOS at the end of May 2025, setting up the transfer of ownership to ATASAM and starting the warranty period of the observatory. Extensive vibration analysis tests were undertaken by ATASAM on the telescope and the Nasmyth platform (and optical table) through most of June 2025. At the same time as KORAY derotator on-sky alignment campaign was initiated by ATASAM – but suffering from delays associated with detrimental weather conditions – the PLACID team decided to proceed to start AIV activities ahead of the summer season, notably to cable up the instrument active components (motorized actuators, SLM) and check functional integrity sufficiently ahead of the commissioning runs planned later in 2025. The rationale was two-fold: taking the opportunity to cable the PLACID electronics and server cabinet infrastructure contemporaneously with the TROIA XAO components, and to reserve enough time to order replacement components (e.g. stepper motors) in case of failure(s), ahead of optical alignment and commissioning activities. Indeed, the last time the PLACID instrument was actively operating in Factory Acceptance configuration dated back to May of 2023 in Yverdon, Switzerland, nearly two years prior, and most active components lifetime already exceed the corresponding warranty period as of July 2025.

This second AIV campaign for PLACID took place in early July of 2025. As illustrated in Figure 6, the PLACID team proceeded to cable up all motor controllers to the PLACID main power box (located under the optical table – not shown), with the latter using a cable tunnel to connect to the Nasmyth server cabinet. The PLACID server rack components (a 1U server unit, a 1U KVM unit, and a 2U interface box) was installed with support from the on-site ATASAM IT team (Figure 7). All active PLACID components – including the customized NIR SLM panel (Figure 8) – were successfully confirmed to operate nominally through the PLACID GUI running on the PC server unit (Figure 9-Left). Additionally, the PLACID team was able to test remote operation of the instrument from the (temporary) control room in the main observatory building downstairs in a closely realistic configuration, using remote desktop connection on the local telescope network (Figure 9-Right). The ability to power cycle the whole instrument (electronics and server rack units) remotely with a network power switch was also successfully tested. The PLACID SLM was removed and packed securely at the end of this campaign, ahead of the optical alignment and internal light source commissioning work, corresponding to the next AIV step foreseen for the instrument.

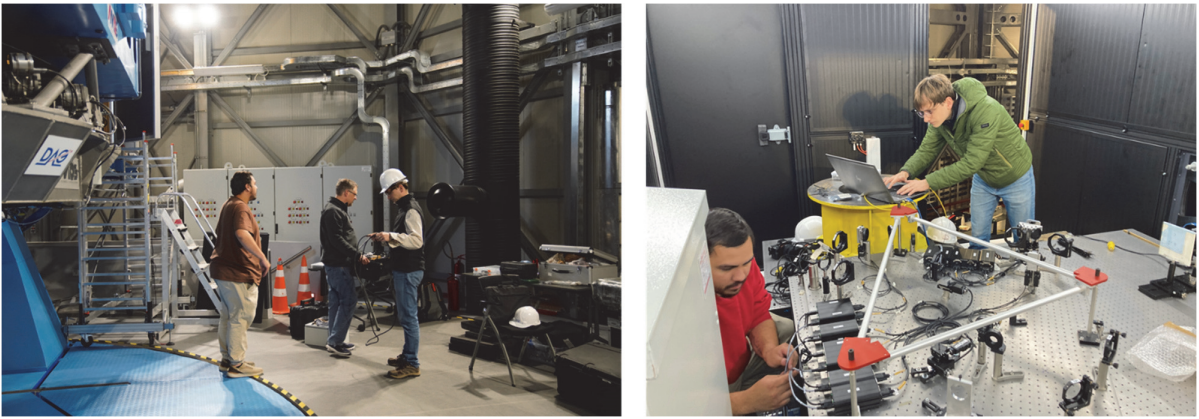

Figure 6. The PLACID team from the University of Bern started the on-site cabling and interfacing work during the early July 2025 campaign.

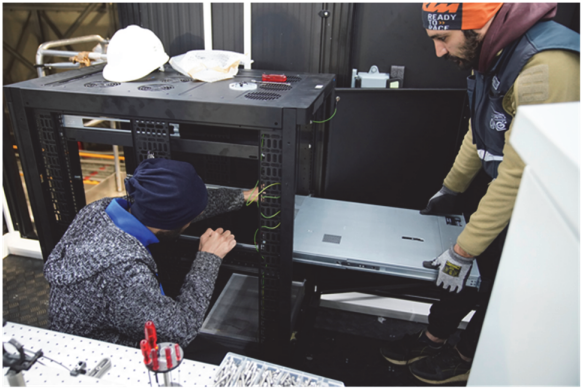
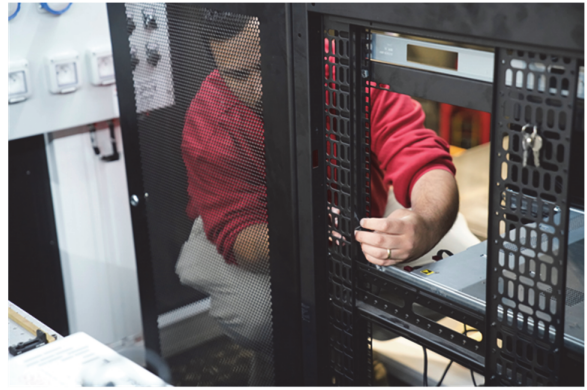
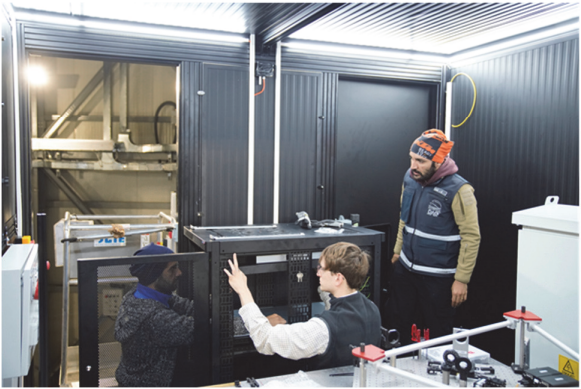
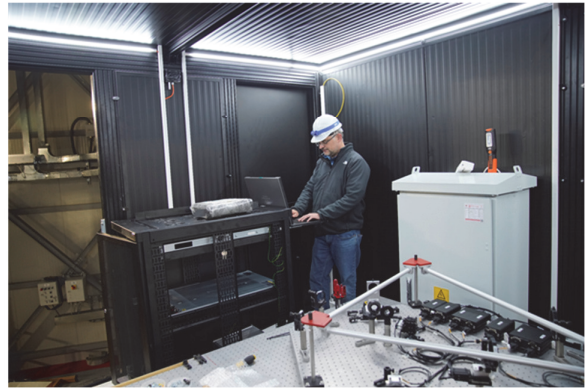

Figure 7. The PLACID cabinet infrastructure – a PC server unit (1U), a KVM unit (1U) and an interface box (2U) - was installed on the Nasmyth server rack on July 3rd, 2025. Note that the depicted cabinet will be later replaced by a liquid-cooled unit over the fall of 2025, to mitigate turbulence inside the Nasmyth enclosure.

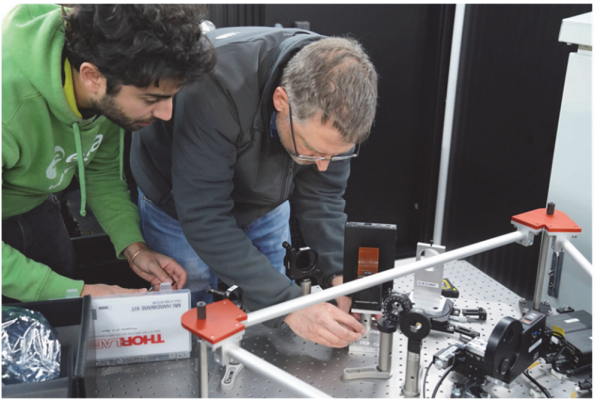
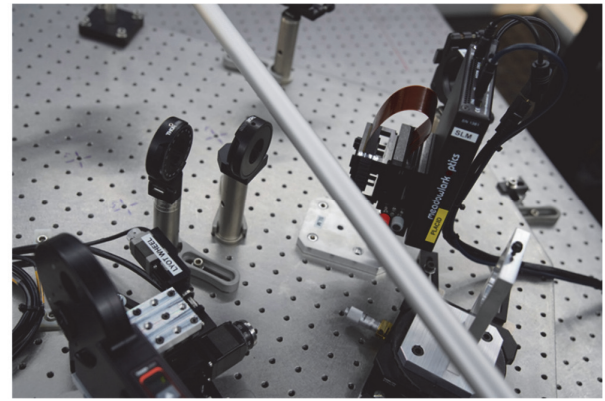

Figure 8. In order to test functionalities of all PLACID active components, the customized H+K-band SLM panel from Meadowlark Corp. was temporarily mounted on the instrument during the July 2025 campaign. Please note the pair of linear polarizer upstream and downstream of the SLM on the right figure.

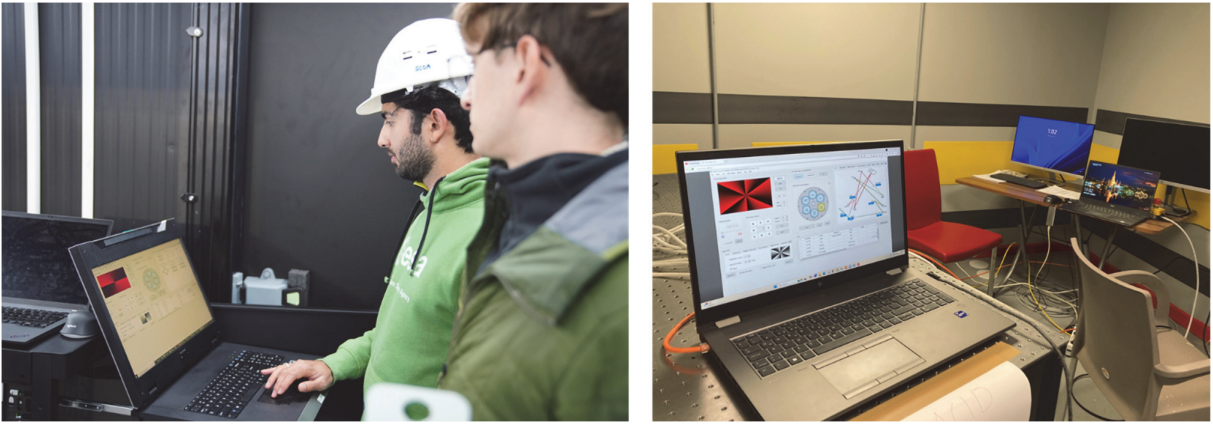

Figure 9. (Left) The PLACID SLM and all motorized components were successfully tested for functional integrity using the control software (GUI) running on the PLACID PC server unit installed in the Nasmyth cabinet; (Right) The same procedure was successfully executed remotely from the DAG (temporary) control room on the observatory floor, included remote power cycling of all sub-systems using a network power switch (not shown).

## 4. DISCUSSION AND OUTLOOK

The upcoming few months of August, September and October 2025 will be mainly dedicated to completing the KORAY derotator on-sky alignment (with respect to the telescope M3 axis) and the opto-mechanical alignment of the TROIA XAO system, by the HEIG-VD and ATASAM teams. Once TROIA is optically aligned with the telescope beam and its own internal stimulus source, it is foreseen that the PLACID optical alignment – and commissioning activities with its own calibration light source - can take place in parallel to the TROIA efforts dedicated to closing the AO loop on-sky. Hence it is currently foreseen that the next PLACID AIV campaign will take place on the course of October of 2025, when the optical alignment work will commence. The contractual period for commissioning PLACID on the Nasmyth with its own calibration source is 60 days, hence it is well possible that the instrument can have first light by the end of 2025, or around early 2026. Assuming no major roadblock occurs on the PLACID side, the exact time window for first light will then mostly depend on TROIA readiness and performance on-sky, as well as on weather conditions at the site. The latter can be relatively random, but the fact that the calendar described above coincides with the winter season 2025-2026 can be detrimental, and could cause delays. In addition, the month of February 2026 will be entirely dedicated to the installation and commissioning of the DIRAC HAWAII-1RG detector by the Australian team (AAO-MacQuarie), hence any first light and on-sky trials with TROIA and PLACID need to happen either by January 2026, or be postponed to March 2026 and beyond. This also means that any on-sky imaging in the near-infrared (H-band) in the time period leading to February 2026 will be performed either with PLACID engineering camera (FLI C-RED 3), with an available FLI C-RED 2 ER currently used in the laboratory in Bern. It is therefore realistic at this point in time to forecast that on-sky commissioning of PLACID will take most of the remaining first half of 2026.

Once its on-sky commissioning will be completed later in 2026, PLACID will the first ever and sole available active programmable "adaptive" coronagraphic high-contrast imaging platform dedicated to exoplanet imaging. Beyond re-imaging of known sub-stellar companions and disks to validate the capabilities of the instrument on-sky, PLACID will be able to address a niche science discovery space around young and compact binary or triple star systems [5,11]. The PLACID team is currently developing a specific branch of the instrument GUI for ADI active imaging of binary stars, to be able to synchronize the DAG telescope control system (TCS) telemetry with the SLM multi-star coronagraphic FPM at its native refresh rate of 30 Hz. The Roddier and Roddier phase mask [23] is currently foreseen as the baseline FPM for this mode, thanks to its straightforward implementation and excellent compatibility with a centrally-obstructed telescope pupil. Additional work planned for 2025 includes the implementation of the phase-shifting variant of the ZELDA WFS for daytime or regular calibration of NCPAs, as well as commissioning of the Ks-band mode of PLACID on a best-effort basis (not contractual). As a versatile XAO-equipped high-contrast imaging facility on a 4-m telescope in the Northern hemisphere, PLACID will likely represent a competitive facility for follow-ups of photometric TESS and PLATO candidates, as well as GAIA DR4 and DR5 astrometric targets of interest. Those aspects align with one of the key missions of the PLACID project, which is to maximize the collaborative, mentoring and nurturing impact for the growing Turkish astronomical community.

Looking beyond the 2026-2027 horizon, PLACID is foreseen to be upgraded with an optional fast imaging mode, fielding a high-speed (~400 Hz) NIR SLM and a ~1.5 kHz C-RED One eAPD camera from First Light Imaging, in the context of the ERC Rapid Active Coronagraphic Exoplanet imaging from a Ground-based Observatory (RACE-GO) project. RACE-GO aims at validating and exploiting on-sky the concepts of coherent differential imaging (CDI), at a speed exceeding the atmospheric coherent time (~10 ms in the near-infrared), in order to freeze the residual post-adaptive optics speckles. This novel observing mode will be relying on the SLM to introduce time-domain synchronous phase diversity in the focal-plane, to dynamically disentangle between coherent stellar speckles and incoherent off-axis astrophysical sources [14]. It is however worth noting that the CDI imaging mode of RACE-GO will be tested on-sky well ahead of time with PLACID and the DIRAC imager, but at much slower speed (0.1 – 1 Hz), allowing to validate CDI performance when dealing with NCPA ("slow speckles"). Overall, also owing the ease of programming any coronagraphic or WFS phase pattern, the PLACID facility will represent a compelling on-sky R&D platform for the HCI community, especially in the context of the ELTs - ELT/PCS in particular - and NASA's Habitable World Observatory (HWO) space telescope.

## ACKNOWLEDGEMENTS

The PLACID and RACE-GO projects have received funding from the Swiss State Secretariat for Education, Research, and Innovation (SERI) as a SERI-Funded ERC 2021 Consolidator Grant, project RACE-GO # M822.00084, following the discontinued participation of Switzerland to Horizon Europe. Part of this work has been carried out within the framework of the National Centre of Competence in Research PlanetS supported by the Swiss National Science Foundation under grants 51NF40 182901 and 51NF40 205606.